\documentclass[aps,10pt,pra,twocolumn,groupedaddress, floatfix, superscriptaddress,showpacs, showkeys,amsfonts]{revtex4-2}
\usepackage{amsfonts}
\usepackage{amsmath}
\usepackage{amssymb, setspace}
\usepackage{graphics,graphicx}
\graphicspath{{./}}
\usepackage{epsf}
\usepackage{subfigure}
\usepackage[%
  colorlinks=true,
  urlcolor=blue,
  linkcolor=blue,
  citecolor=blue
]{hyperref}
\usepackage[all]{hypcap}
\usepackage{color}
\usepackage[utf8]{inputenc}
\usepackage{dsfont}
\usepackage{enumitem}
\usepackage[utf8]{inputenc}

\usepackage{comment}
\usepackage{orcidlink}

\usepackage{natbib}
\usepackage{braket}
\usepackage{physics}

\usepackage[normalem]{ulem}
\usepackage{xcolor}

\newcommand{\be}{\begin{equation}}
\newcommand{\ee}{\end{equation}}
\newcommand{\bea}{\begin{eqnarray}}
\newcommand{\eea}{\end{eqnarray}}
\newcommand{\eq}[1]{Eq.~\eqref{#1}}

\newcommand{\eqss}[2]{Eqs.~\eqref{#1}-\eqref{#2}}

\newcommand{\fig}[1]{Fig.~\ref{#1}}
\newcommand{\figs}[2]{Figs.~\ref{#1}-\ref{#2}}
\newcommand{\bem}{\begin{multline}}
\newcommand{\eem}{\end{multline}}

\newcommand\identity{1\kern-0.25em\text{l}}

\newcommand{\apporsm}[1]{App.~\ref{#1}}

\begin{document}

\title{Feedforward suppression of readout-induced faults in quantum error correction} 

\newcommand{\ibmil}{IBM Quantum, IBM Research -- Israel, Mount Carmel, Haifa 31905, Israel}
\newcommand{\ibmny}{IBM Quantum, T.~J.~Watson Research Center, Yorktown Heights, New York 10598, USA}

\author{Liran Shirizly\orcidlink{0009-0002-4597-0126}}
\email{liran.shirizly@ibm.com}
\affiliation{\ibmil}
\author{Dekel Meirom\orcidlink{0000-0002-8500-3838}}
\email{dekel.meirom@ibm.com}
\affiliation{\ibmil}
\author{Malcolm Carroll}
\email{malcolm.carroll@ibm.com}
\affiliation{\ibmny}
\author{Haggai Landa}
\email{haggai.landa@ibm.com}
\affiliation{\ibmil}

\begin{abstract}
    Qubit measurements in quantum devices involve various types of errors, including erroneous state determination, correlated preparation errors and measurement-induced leakage from the computational states. We propose a feedforward protocol to reduce readout-induced faults, applicable for qubits with errors biased between the different states, in settings like quantum error correction with repeated measurement cycles. The method consists of an adaptive readout sequence conditioned on each check qubit's readout result from the previous cycle, which is optimized for the expected measured state. Focusing on a simple realization of conditionally flipping (by an X gate) the state of check qubits before their measurement, we investigate the effect of such state-dependent errors using simulations in the setup of a low-density parity check code. We show that the suggested protocol can reduce both logical errors and decoding time, two important aspects of fault-tolerant quantum computations.
\end{abstract}

\maketitle

Components of quantum error correction (QEC) and fault-tolerant computing are being realized with various platforms \cite{andersen2020repeated, satzinger2021realizing, sundaresan2023demonstrating, PhysRevX.11.041058,krinner2022realizing, postler2022demonstration,google2023suppressing, gupta2024encoding, bluvstein2023logical, google2025quantum, putterman2025hardware, reichardt2024logical}. A dominant approach in the field is based on quantum circuits realizing stabilizer codes, wherein errors in data qubits (which store the quantum information) are repeatedly mapped to check qubits that are then measured to extract the error syndromes \cite{RevModPhys.87.307}. A measurement-based scheme \cite{gottesman2014fault, breuckmann2021quantum} also forms the basis of efficient implementations of logical gates on state-of-the-art quantum low-density parity check (LDPC) codes \cite{cohen2022low,cross2024improved, williamson2024low, swaroop2024universal,cowtan2025parallel,zhang2025time}. Such codes therefore share a significant reliance on the performance of mid-circuit measurements of check qubits. 

Measurement results of check qubits constitute syndromes that are fed to a decoder whose task is to determine the changes in the state of the data qubits, and specify a possible correction procedure. Instead of applying an actual correction, the Pauli frame (including the  logical one) can be tracked by the decoder \cite{knill2005quantum, Chamberland2018faulttolerant}.
Pauli frame tracking is often required since decoding is a computationally intensive task, while solid-state qubits, for example, operate at relatively fast clock times \cite{kjaergaard2020superconducting,RevModPhys.95.025003}. Approaches to reduce the real-time decoding computational burden are therefore an active area of research from both algorithmic and hardware perspectives \cite{roffe_decoding_2020, liyanage2023scalable,bombin2023modular, caune2024demonstrating, gong2024toward,chen2025improved, jones2024improved,kaufmann2025blockbp, demarti2024decoding, higgott2025sparse, PhysRevLett.133.240602,ott2025decision, beni2025tesseract}. 

\begin{figure}
\centering
\includegraphics[width=0.48\textwidth]{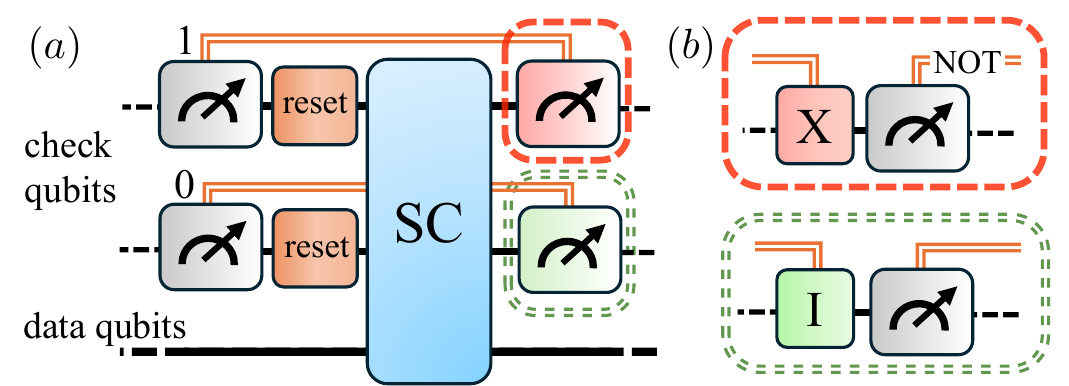}
\caption{
A schematic depiction of the adaptive readout sequence, based on the outcome of each qubit's previous measurement. (a) As an example focused on two check qubits of a stabilizer-based error correction code, following the measurement step shown on the left and in the absence of any further error, the first and second check qubits are again expected to be measured in the states $\ket{1}$ and $\ket{0}$ respectively (at the end of the syndrome cycle ``SC''). 
The measurement type (indicated by color) applied to each qubit is conditioned and adapted to its expected state. (b)
A possible realization of the adapted measurement is a conditional flip (an X gate) applied to each qubit that is expected to be in $\ket{1}$; in this example, setting the first qubit back to $\ket{0}$ while the second one is left untouched. A classical NOT on the first qubit's measured bit would recover the expected syndrome at the end.\label{fig:scheme}}
\end{figure}

In typical implementations, a check qubit's value will tend to repeat itself at the end of each syndrome cycle until a fault leads to a flip. The fraction of check qubits measured in the excited state is expected to saturate at one half over many syndrome cycles when using Pauli frame tracking in a system with a computational basis formed by an effective ground state and excited state. For example, in Calderbank, Shor, Steane (CSS)-type codes, this is the case for X checks starting a cycle with all qubits in the ground state due to random collapse, and for the Z checks this follows from a random walk on the syndromes. 

With many qubit modalities, error rates are asymmetric between the computational $|0\rangle$ and $|1\rangle$ states during measurement and other gate operations. 
Importantly, leakage of the qubit state from the computational space (or qubit loss) is a significant issue, and readout-induced leakage may be stronger from one particular state, often the excited state \cite{PhysRevLett.117.190503, PhysRevApplied.20.054008,nesterov2024measurement,dumas2024measurement, PhysRevA.111.012611,PhysRevLett.134.100601, bista2025readout}.
Choices in the implementation of QEC could be customized to minimize exposure to the higher error rate states, significantly improving performance, and complementing various hardware approaches to enhance active and passive leakage reduction  \cite{PRXQuantum.2.030314,mcewen2021removing,miao2023overcoming, lacroix2025fast,kim2025fast,thorbeck2024high,PhysRevLett.130.250602,camps2024leakage,yang2024coupler,fowler2013coping,ghosh2015leakage,brown2019leakage,suchara2015leakage}.

In the domain of readout-error mitigation efforts are centered on increasing accuracy given a large number of shots \cite{PRXQuantum.2.040326,PhysRevA.106.012423,doi:10.1126/sciadv.abi8009,PRXQuantum.5.030315,PRXQuantum.6.010307, henao2023adaptive,koh2024readout,trex}, with Ref.~\cite{PhysRevA.103.022407} improving sampling results with X gates added before the readout of qubits that are likely to be in the state with a higher error at the termination of the circuit. 
Those approaches are not necessarily practical for QEC and single-shot improvements are required.
Current works addressing such improvements use additional qubits, which complicates the required device connectivity \cite{PhysRevA.105.012419,PhysRevLett.134.080801}.
Therefore, new methods for error suppression are needed.

In this paper we propose an adaptive gate approach to reduce faults in the error-prone state during mid-circuit readouts that are repeated as part of a series of identical circuits, e.g., the syndrome cycles in a QEC setup. The protocol is designed to suppress the underlying physical cause of readout, preparation and leakage errors by using dynamic circuits with a feedfoward (conditional) operation, which naturally fits within a Pauli frame-tracking scenario. In addition to the reduction in logical error rates, we find that practical decoding algorithms execute faster as fewer detectors are unsatisfied.

{\it The adaptive readout method.} 
The essence of the method, as schematically depicted in \fig{fig:scheme}(a), is in adapting the measurement gate of each check qubit to its expected state at the end of every QEC syndrome cycle. The adapted measurement gate sequence is selected based on the readout result from the previous cycle. More generally, the conditional logic could be extended to depend on a majority vote or some other processing of the last $m$ readout results. The state-adapted measurement gate sequence can include, e.g., tuning the readout pulse parameters (amplitude, frequency, and other hardware specific parameters), or the discrimination boundaries and algorithm differentiating between the qubit states, allowing an overall optimization of the error rate. 
Such adaptations can be advantageous in particular for competing errors (e.g., when leakage is stronger from one qubit state while readout error is higher in the other state). Similarly, these adaptations could potentially be used to reduce some measurement-induced errors from one state at the cost of increasing others (e.g., decreasing leakage while increasing readout errors), even if these errors are not biased originally, as discussed in \apporsm{Sec:AdaptReadout}.

In the following we focus on the adaptation by insertion of a conditional gate before the measurement pulse. Without loss of generality, we choose that the excited state $|1\rangle$ suffers from higher rates of readout-induced faults than the ground state $|0\rangle$. Readout is restricted to this basis and measurement of other axes is facilitated by single qubit operations (e.g., Hadamard gates) before the measurement.
Then a conditional local X gate can be applied to each check qubit based on its last readout bit, as in \fig{fig:scheme}(b). 
Check qubits that have been correctly measured in the state $|1\rangle$ in the previous syndrome cycle, are now expected to be measured in the ground state $|0\rangle$ when applying the adaptive gate before the measurement (i.e., without the adaptive gate and assuming Pauli frame tracking, the check would otherwise be again mapped to $|1\rangle$). A classical bit flip is applied to the measurement outcome, keeping the original meaning for the syndrome as if there was no adaptive gate. The reduced frequency of measured $|1\rangle$ is then expected to reduce the overall number of errors.
Qubits whose measurement result was 1 due to a readout error will be flipped erroneously, but this will not propagate beyond one more cycle. 
In a syndrome cycle implementation that avoids resets of the check qubits, the protocol should be conditioned on the previous two measurements -- applied if $0$ was measured in the previous cycle and $1$ was measured in the one before it.

This method can be applied to various error correction codes in order to suppress readout-induced faults. In the rest of this work we focus in detail on a specific code, the LDPC Gross code \cite{bravyi2024high}, where we provide numerical evidence indicating that this protocol could reduce logical error rates and decoding times, in particular accounting for leakage errors. 

{\it Decoding model.} 
We simulate the LDPC Gross code with 144 data qubits and the same number of check qubits (encoding 12 logical qubits), which is described in detail in \cite{bravyi2024high}. The Gross code is a CSS code and X-type and Z-type syndromes are decoded separately.  
The decoding simulations are based on a standard circuit-level noise model as implemented in \cite{bravyidecoder}. We execute Monte Carlo simulations of repeated applications of the syndrome cycle together with random realizations of Pauli errors with prescribed probabilities following every gate. We focus on error parameters consistent with near-term state-of-the-art superconducting qubits, and our goal in the numerical simulations is to explore the overall trends in the parameter space, which necessarily come with some tradeoff between various choices. 
We must stay within the boundaries of stochastic noise (very close to a Pauli model \cite{google2023suppressing}) as it remains the only feasible option at those scales despite its known limitations \cite{greenbaum2017modeling,iverson2020coherence,PhysRevA.111.022613, malekakhlagh2025efficient}. 

The syndrome cycle is a Clifford circuit, which can be thought of as a logical identity circuit, composed of single-qubit gates of different types and two-qubit CNOT gates coupling data and check qubits. The single-qubit gates include preparation (qubit initialization) and readout (measurement) gates of the check qubits, and some idle (identity) gates on the data qubits. The standard noise model includes errors in the preparation and the readout (wherein an error is a state flip), in the idle gates (a depolarizing single-qubit Pauli channel), and in the CNOT gate (a depolarizing two-qubit Pauli channel).

We extend this model with state-dependent errors of readout, readout-induced leakage, and further relevant error types as detailed below. 
To account for readout-induced leakage, we consider a simplified model in which only check qubits in the state $|1\rangle$ can leak outside of the computational levels, and leakage is calculated only during readout (as readout-induced faults is the focus of this work). A leaked qubit always reads 1 and its state does not change during the CNOT gates of the syndrome cycle. Within this model, a CNOT gate between a data qubit and a leaked check qubit has some probability of incurring (independently) a Pauli Z or X error to the data qubit, which we denote as the backaction error rate. This is a plausible reduction of the standard two-qubit Pauli error.
A leaked qubit has also a seepage rate, which is its probability to return to the state $|1\rangle$ during one syndrome cycle. In our simplified model the seepage is calculated only at the end of each cycle, during readout gate. This is an effective averaged description of spontaneous and induced processes that take the qubit back to the computational space with a prescribed mean rate.

We define $p(a|b)$ to be the error rate of measuring $a$ when the qubit should have measured $b$, and parameterize the qubit readout errors according to the average readout error and the bias,
\be p_{\rm read} = \frac{1}{2}\left[p(0|1) + p(1|0)\right], \quad p_{\rm bias} = \frac{1}{2}\left[p(0|1) - p(1|0)\right],\label{Eq:p_r_b}\ee
To simulate state-dependent errors with different rates as in \eq{Eq:p_r_b}, we start every simulation with a random initial state of the data qubits, in order to induce a random Pauli frame. Moreover, since the competing leakage and seepage (with different rates) lead to a steady state in the population of leaked qubits, we first study its characteristics (see \apporsm{Sec:LeakEvolution}), and in the following we start each simulation with a random initialization of leaked qubits corresponding to this mean population.

In addition, we consider two types of idle (identity) gates on data qubits; short, simultaneous with a CNOT gate (denoted $p_{\rm id,\,s}$), and long ($p_{\rm id,\,l}$, with an increased associated error), simultaneous with the readout gates on the check qubits. 
We use uniform qubit error rates, fixing the preparation, short idle and CNOT error parameters (see Tab.~\ref{parameters_table} for a summary of all parameters);
\be p_{\rm prep} = 0.001, \quad p_{\rm id,\,s} = 0.001, \quad p_{\rm cnot} = 0.001,\label{eq:fixed_rates}\ee
and the state-dependent readout parameters,
\be p_{\rm read} = 0.02,\quad p_{\rm bias} = 0.005 .\label{eq:readout_rates}\ee
The gate parameters reflect values expected to be achieved in early systems running large-scale QEC, while the readout error rates are on the higher side of mean values in large superconducting quantum processors, as the mid-circuit measurements are expected to be faster and more error prone than the reported terminal readouts. As exemplified by our results in the following, it may be advantageous to accommodate higher readout errors in order to lower other significant readout-induced faults (especially leakage or the data qubits' idle time).
In the data below we vary the leakage, seepage, backaction and long idle error rates.
In the simulation of the qubit flip protocol, we test the effect of an increased long-idle error (representing a tradeoff accounting for the longer duration due to the X gate and possibly some delays due to the classical conditioning), and we also account for a flip error in the X gate.

Leakage rates on the order of $10^{-3}$ have been reported with transmon qubits \cite{sundaresan2023demonstrating} and we vary the leakage rate in this regime. 
Regarding the seepage rate, in the simulations it is immediately clear (and has been known experimentally) that if leaked qubits are not reset to the computational space fast enough, a leaked population quickly builds up to make error correction impractical. Therefore some form of leakage removal must be assumed, and we vary seepage rates in the regime of order 1. A high seepage rate (corresponding to a frequent application of leakage removal throughout the device) must clearly incorporate some cost to it (in operation time at least), a tradeoff that we explore below.

The logical error rate is also very sensitive to errors on the data qubits being induced by leaked check qubits during CNOT gates. This type of error is highly dependent on the details of qubit and gate operations, however, we see that if this backaction probability is taken naively to be of order 1 (corresponding to gates involving leaked qubits being blindly executed), the logical error rates rise to a detrimental level. Therefore, this scenario is uninteresting to simulate, and we focus on the regime in which this probability is a few percent, which may reflect qubit design or can be thought of as modeling mitigation schemes that assume identifying the leaked qubits (with a high probability) in real time \cite{bultink2020protecting,varbanov2020leakage,kanazawa2023qutrit,mude2025efficient}.
It is expected and confirmed that the higher the leakage and backaction error rates are, and the lower the seepage rate is, the more beneficial the flip protocol will be.

{\it Results.} The data presented in the rest of this paper is based on simulating multiple ``shots'' (independent simulation runs) for each combination of parameters, decoding windows of $N=12$ syndrome cycles starting with the belief propagation (BP) algorithm \cite{roffe_decoding_2020}. BP is an iterative approach to syndrome decoding, and the mean executed number of iterations is observed to grow with the complexity of the error. A cutoff of a maximal number of iterations is imposed at 200 iterations. If BP does not converge, a secondary decoder (with a significantly higher algorithmic complexity) is invoked, which is the ordered statistics decoding (OSD) algorithm of order 7 with combination sweep as in \cite{ Roffe_LDPC_Python_tools_2022,bravyi2024high}. The decoder is based on a linearization of the noise model, searching for the (estimated) most likely linear combination of the syndromes of single faults, with state-independent error probabilities defined by \eqss{eq:fixed_rates}{eq:readout_rates} together with the long-idle error rate $p_{\rm id,l}=0.005$. For $N$ syndrome cycles in a run of the simulation, if $p_N$ is the probability of a logical error on one or more of the encoded qubits, the logical error per syndrome cycle is customarily \cite{bravyi2024high} defined as $1-(1-p_N)^{1/N}\approx p_N/N$, and is reported below together with other decoding characteristics.

\begin{figure}
\centering
\includegraphics[width=0.48\textwidth]
{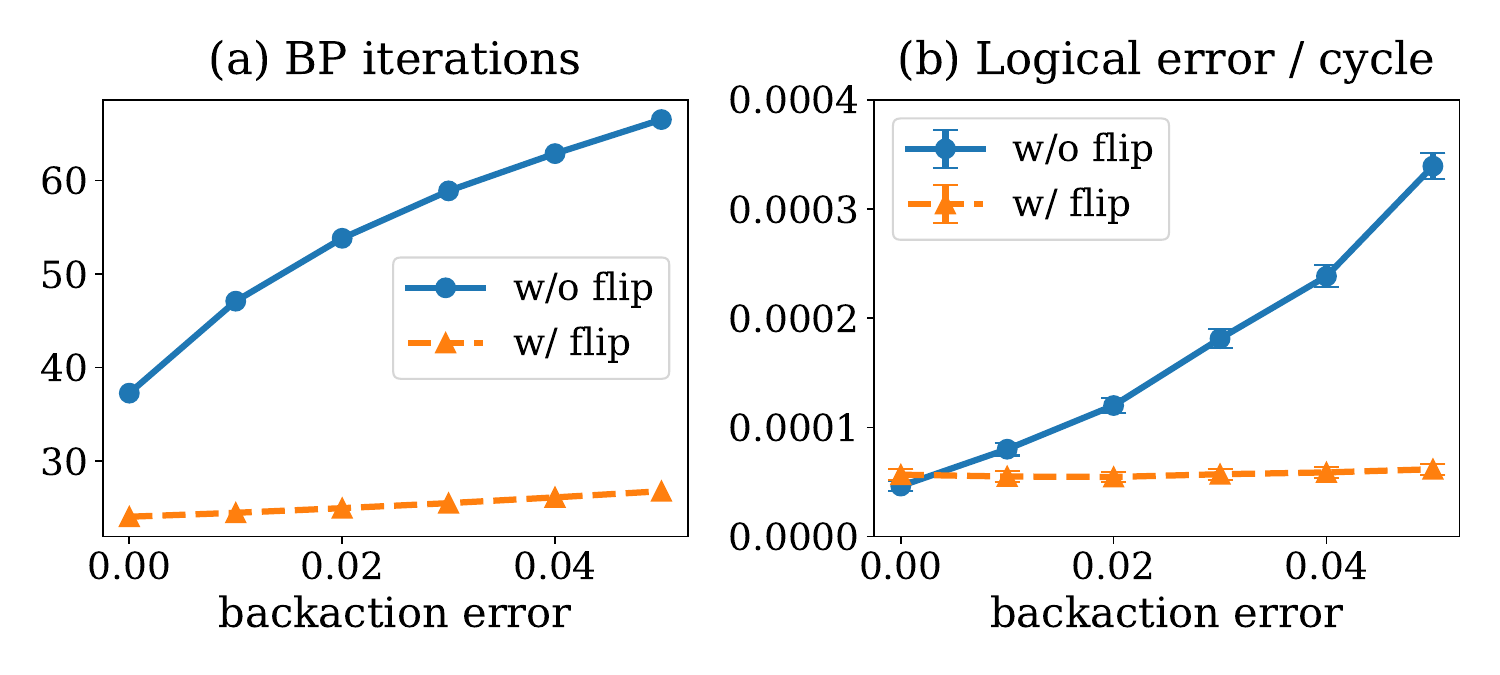}
\caption{Results from decoding of simulated syndrome measurements as a function of the backaction error probability (corresponding to leaked check qubits inducing independent Z and X errors on the coupled data qubits), with the data comparing the flip protocol (denoted as ``w/ flip'') vs.~its absence (denoted as ``w/o flip''), for error rates as in \eqss{eq:fixed_rates}{eq:flip_rates}. (a) The mean number of BP iterations in each decoding execution. The decoding time is significantly reduced with the flip protocol due to fewer BP iterations being needed. (b) The logical error rate per syndrome cycle, showing a significant reduction to a nearly constant level. See the text for a detailed discussion.
\label{fig:back}}
\end{figure}

For generating faults in the syndromes decoded for the data of \fig{fig:back} we fix the following parameters
\be p_{\rm id,\,l} = 0.005, \quad p_{\rm leak} = 0.002,\quad p_{\rm seep} = 0.2,\label{eq:medium_rates}\ee
and scan the magnitude of the backaction error across a small range (in the regime of a few percent, as explained previously).
For generating syndromes with the flip protocol we increase the long idle gate error, and add also the flip error parameter;
\be \quad p_{\rm id,\,l} = 0.006,\quad p_{\rm flip} = 0.0002. \label{eq:flip_rates}\ee
The first data point in \fig{fig:back}, with no backaction error, shows that if all that leakage does is to make a check always read 1, the decoder is able to handle well the leaked checks and the bias in readout error, resulting in no improvement of the logical error rate due to the flip protocol with the above error rates (and even a slight degradation with the modeled increased of $p_{\rm id,l}$). 

\begin{figure}
\centering
\includegraphics[width=0.48\textwidth]
{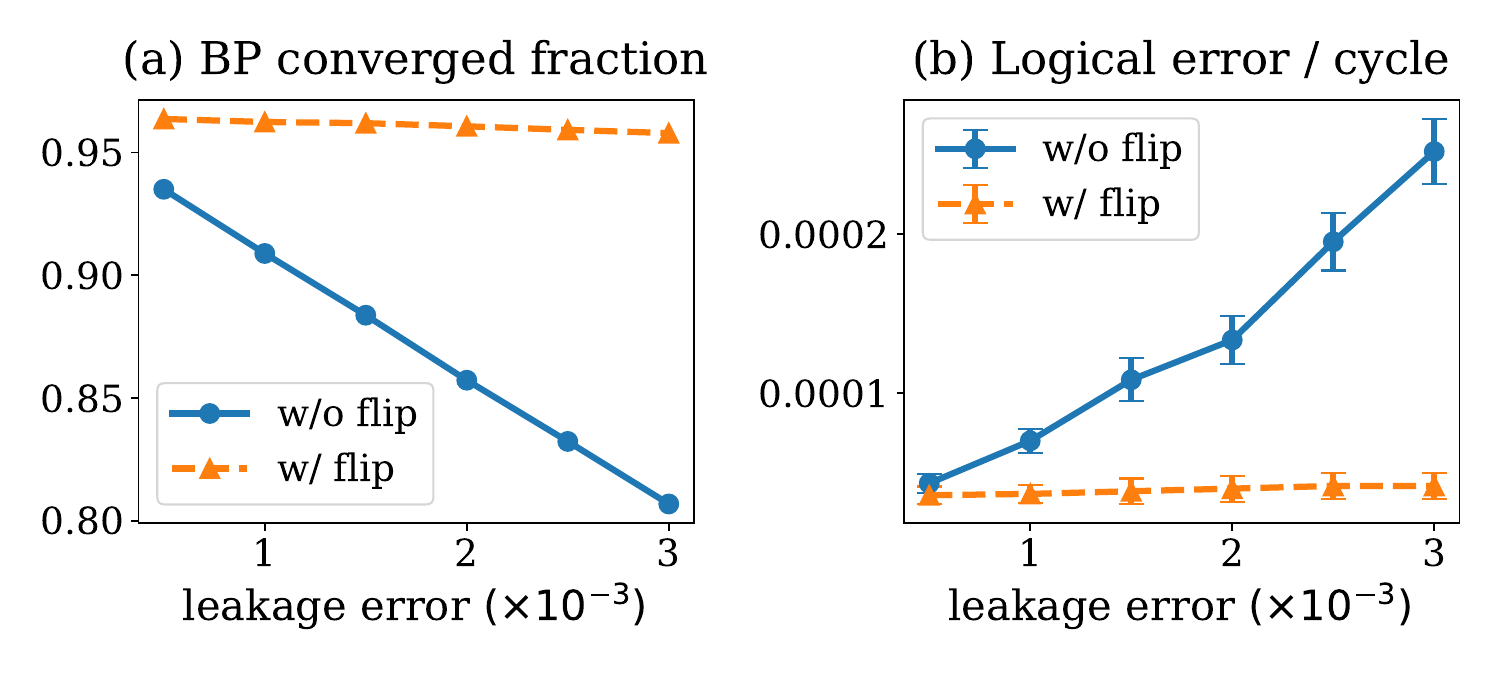}
\caption{Decoding of simulated syndrome measurement circuits as a function of the leakage rate, comparing the flip protocol vs.~its absence, with error rates in \eqss{eq:fixed_rates}{eq:leak_rates}. (a) The fraction of BP attempts that converged. The increased convergence seen with the flip protocol dramatically reduces the decoding time by eliminating the need for secondary decoder calls. (b) The logical error rate per syndrome cycle, showing again a significant reduction with the flip.\label{fig:leak}}
\end{figure}

Nevertheless, even in that scenario, the flip protocol would improve significantly the decoding time, an improvement that becomes even more pronounced for increased error rates. This is seen in the reduction in the mean number of BP iterations seen in \fig{fig:back} and furthermore, because of the fraction of converged iterations increasing significantly with the flip protocol [not directly shown in \fig{fig:back} but a similar behavior can be seen in \fig{fig:leak} described below]. This is consistent with the large increase in the mean ground-state population of check qubits just before readout, which rises from 50\% to approximately 93\% for all data points, and the related population of leaked check qubits just before readout (see \apporsm{Sec:LeakEvolution}). The number of BP iterations is plausibly reduced due to less detectors being unsatisfied, and the better convergence of BP leads to a large reduction of decoding time following the fewer (costly) secondary decoders executions.

A related parameter scan is presented in \fig{fig:leak} where we take an example of an even higher seepage rate 
\be p_{\rm seep} = 0.33, \quad p_{\rm back} = 0.05,\label{eq:leak_rates}\ee
while the backaction probability is fixed to the maximal value shown in the scan of \fig{fig:back}.
Scanning the leakage rate in a relevant range we see again an expected sensitivity in the logical error rate to the leakage probability, and a strong suppression of the error with the flip protocol. We note again that here as well, even at very low leakage, the flip protocol reduces the expected decoding time, with a lower mean number of BP iterations and a higher fraction of converged ones.

\begin{figure}
\centering
\includegraphics[width=0.48\textwidth]
{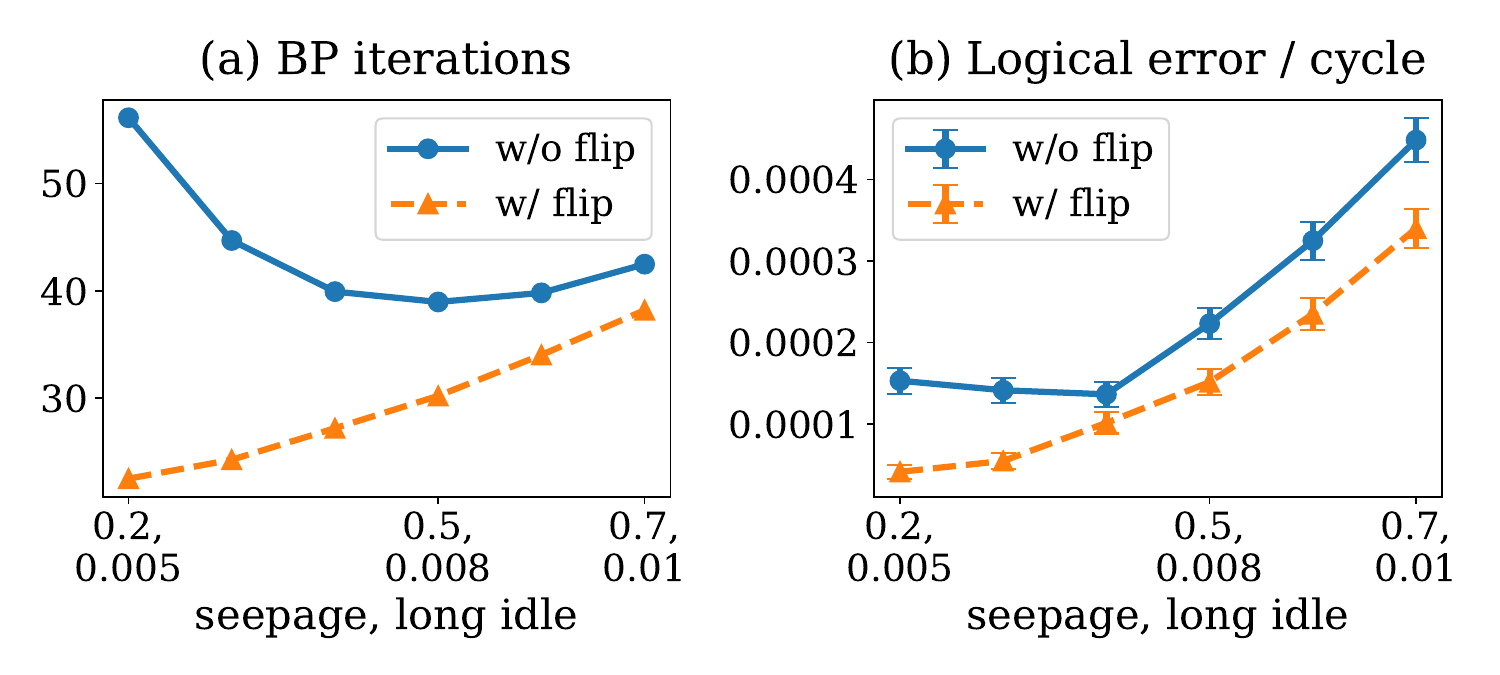}
\caption{Decoding results when varying the seepage rate (between 0.2 and 0.7 in increments of magnitude 0.1) simultaneously with the error in the long-idle gate (between 0.005 and 0.01 in steps of 0.001), with the other error rates as in \eqss{eq:fixed_rates}{eq:readout_rates}. (a) The mean number of BP iterations. (b) The logical error per syndrome cycle. Modeling a tradeoff between the efficacy of a leakage removal process and its cost in terms of the prolonged idling of data qubits, the non-monotonic behavior of the decoding quantities without the flip emphasizes the importance of having a variety of tools at hand for optimizing complementing approaches.
\label{fig:id_l}}
\end{figure}
 
In a typical hardware implementation of readout it can be expected that varying the duration of the readout step would incur some tradeoff between different error rates. Specifically, increasing the seepage rate could be expected to require increasing the long idle duration, resulting in an increased error probability on the idling data qubits. To investigate this aspect we show in \fig{fig:id_l} the decoding results when varying together the seepage rate and the long-idle error (neglecting here a separate increase of the idle error when the flip protocol is performed). A non-monotonic behavior of the decoding penalties is observed without a flip protocol, exemplifying how such an approach can serve as a guideline to finding optimal operational parameters. By taking advantage of the flip protocol, the global lowest logical error and decoding time are obtained together for the presented parameters when -- perhaps surprisingly -- the lowest seepage rate is chosen. Combining numerical input with direct experimental optimization of complex tradeoff choices in hardware parameters, software aspects and decoding model can be expected to become increasingly more relevant, encompassing not only logical errors \cite{domokos2024characterization,marton2024optimal} but other computational aspects such as decoding time, as above.

{\it Conclusion and outlook}.
The adaptive readout method the we propose can significantly reduce both logical errors and decoding time. The highest impact of the protocol is expected if the conditional or adapted gates on the check qubits could be implemented without delays. While current hardware supports conditional operation in real time, it requires a finite time to execute after inquiring the measurement outcome. As our protocol doesn't require conditional operations immediately after measurement, in principle the whole duration of a syndrome cycle can be used to process and prepare the required instruction.

Analyzing the expected tradeoffs between different mechanisms for suppression of errors in QEC settings could be a powerful tool to test global optimization approaches (and calibration approaches as in \apporsm{Sec:AdaptReadout}). The simulation code used to produce the data for this work is available as open source \cite{qec_flip_repo}, and it can be extended and used to quantify more elaborate error models or conditional protocols. As another example for state-dependent noise, in \apporsm{Sec:CorrelatedPrep} we probe preparation faults correlated to readout errors (i.e., a preparation error occurring conditioned on a readout error, as in the so-called ``conditional reset'' scheme that benefits from a fast implementation given the preceding measuring), seeing a dramatic increase in the logical error rates under certain conditions, with some reduction using the flip protocol.

The presented approach therefore broadens the palette of cost-benefit options with which to improve decoder performance and better suppress effects like readout error and leakage against modest costs that might include some hardware modifications to enable feedforward, a possible feedforward delay and impact of the additional gate insertion (e.g., idling errors of data qubits and the X gate error). Beyond the directly reduced logical error, the reduction of decoding time could have significant implications, allowing for real-time decoding of larger codes and employing larger check matrices encompassing more faults, eventually further reducing the logical error rates. \\

\bigskip
{\it Acknowledgements.}
We thank Luke Govia, Maika Takita, David McKay, Ted Yoder, Tomas Jochym-O'Connor, Michael Beverland, Andrew Cross, Lev Bishop and Doug McClure for very helpful feedback.
Research by H.L. and L.S. was partially sponsored by the Army Research Office and was accomplished under Grant Number W911NF-21-1-0002. The views and conclusions contained in this document are those of the authors and should not be interpreted as representing the official policies, either expressed or implied, of the Army Research Office or the U.S. Government. The U.S. Government is authorized to reproduce and distribute reprints for Government purposes notwithstanding any copyright notation herein.

Elements of this work are included in a patent application filed by the International Business Machines Corporation with the US Patent and Trademark Office.

\begin{table}[h]
\begin{tabular}{ |p{1.6cm}||p{2.1cm}|p{2.2cm}|p{2.cm}|  }
 \hline
 \multicolumn{4}{|c|}{Parameters and their value in the main figures} \\
 \hline
 Parameter& \fig{fig:back} &\fig{fig:leak}&\fig{fig:id_l}\\
 \hline
 $p_{\rm read}$   & 0.02    &0.02&   0.02\\
 $p_{\rm bias}$&   0.005  & 0.005   &0.005\\
 $p_{\rm prep}$ &0.001 & 0.001&  0.001\\
 $p_{\rm id,\,s}$    &0.001 & 0.001&  0.001\\
 $p_{\rm cnot}$ &   0.001  & 0.001 & 0.001\\
 $p_{\rm id,\,l}$& 0.005 / 0.006  & 0.005 / 0.006   &0.005 - 0.01\\
 $p_{\rm leak}$& 0.002  & 0.0005 - 0.003&0.002\\
 $p_{\rm seep}$& 0.2  & 0.33&0.2 - 0.7\\
 $p_{\rm flip}$& 0.0002  & 0.0002&0.0002\\
 $p_{\rm back}$& 0 - 0.05  & 0.05&0.025\\
 \hline
\end{tabular}
\caption{A summary of all the parameters used in the different simulations presented in \figs{fig:back}{fig:id_l}.}
\label{parameters_table}
\end{table}

\appendix

\section{Optimizing the readout parameters}
\label{Sec:AdaptReadout}
The adaptations of the readout parameters according to the expected measured state can be optimized in some cases, e.g., when there are competing mechanisms of readout-induced faults, or even in the simpler case of biased readout error. To give a concrete example that is simple to analyze, we can focus on readout errors alone and consider a favorable modification of the discrimination. It can happen that for the ``standard'' calibration of the readout pulse, leakage is stronger from one qubit state while the readout error is larger for the other state.
In such a case, the readout gate can be tailored to each state such that the readout errors and induced faults will be reduced (with some tradeoff between different types of such faults). Each adaptation will have some effect and the overall error rates can be optimized. This approach might be beneficial even if the ``standard'' errors were symmetric.

For a rudimentary analysis, we assume a standard calibration of the readout resulted for each qubit in a minimal mean readout error $p_{\rm read}$ and some given value of $p_{\rm bias}$ that depends on the qubit and measurement properties. A simple adaptation of the readout would be to set a different threshold of discrimination, decreasing $p(1|0)$ while paying with $p(0|1)$. Defining the prior probability $p_0$ of a check qubit to be in the 0 state just before readout at the end of a syndrome cycle, the results of \apporsm{Sec:LeakEvolution} show that with the flip protocol we get $p_0\approx 0.93$ (for the parameters described there). Then it is clear that by minimizing $p_0\times p(1|0) + p_1 \times p(0|1)$ in the setting of the QEC syndrome cycle, the overall readout error rate can be reduced. Since $p_0$ depends on all errors during the syndrome cycle, the best optimization would require an experiment applying the protocol to known states and adapting the measurement gates such that a minimal overall error (or decoding time) is observed. 

\section{Preparation errors correlated with a readout error}
\label{Sec:CorrelatedPrep}

\begin{figure}
\centering
\includegraphics[width=0.48\textwidth]
{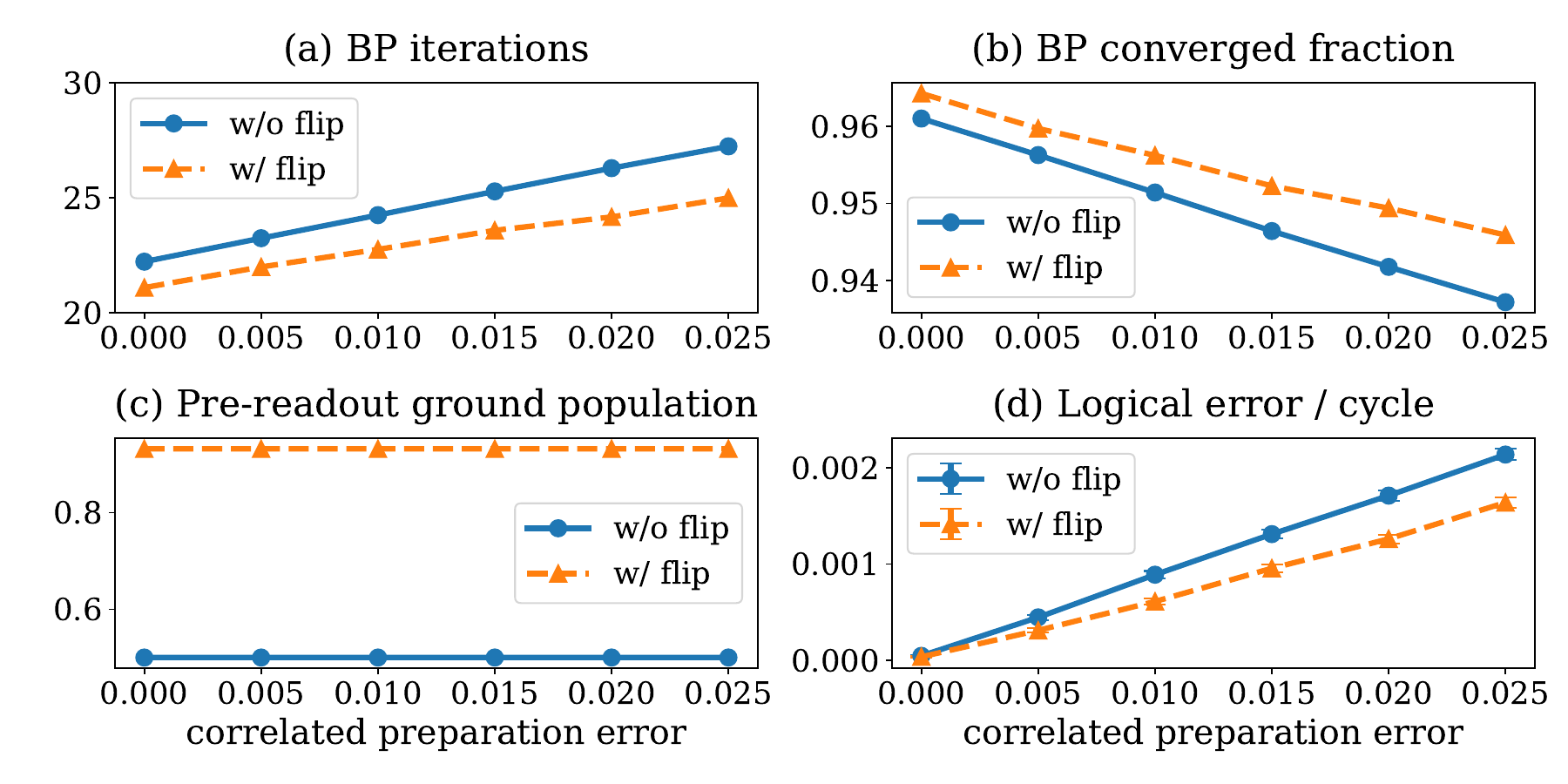}
\caption{The dependence of decoding characteristics on preparation errors occurring with the probability $p_{\rm corr}$ conditioned on a readout error having occurred in the preceding readout step. The error parameters are as in \eqss{eq:fixed_rates2}{eq:readout_rates2} and $p_{\rm id,l}=0.005$. (a) The mean number of BP iterations per decoding window. (b) The fraction of BP algorithm runs that converged. (c) The mean population of check qubits in the ground state just prior to the syndrome cycles' readout step. (d) The logical error rate per cycle. A significant increase in decoding errors is observed for even small residual preparation errors. See the text for a discussion.
\label{fig:corr}}
\end{figure}

Our state-dependent simulations allow us to probe preparation errors correlated to a readout error, modeled as occurring with some probability $p_{\rm corr}$ conditioned on a readout error having occurred in the immediately preceding readout gate. With a high residual preparation error (of order 1 after a readout error) a dramatic increase in the logical error rates is observed. Although this would be the naive scenario for the mid-circuit measurement and reset protocol known as conditional reset (in which the reset is directly based on the readout result), steps for the reduction of this error can be anticipated.

In \fig{fig:corr} we therefore present some simulation results focused on the regime of a residual correlated preparation error of a few percent and no leakage at all. Even for such small residual correlated preparation errors, the logical error could increase by an order of magnitude, and the flip protocol shows only a modest improvement in decoding quantities in the regime of parameters presented here. It is beyond the scope of the current work to explore the error budgets of different suppression approaches for these errors, which can include properly accounting for such errors in the decoder or reducing their probability in the physical implementation, again necessarily accompanied with some tradeoff costs.

\section{Evolution of the population of leaked qubits}\label{Sec:LeakEvolution}

Considering the dynamics of qubit leakage and seepage, starting from an initial state with no leaked qubits, the number of leaked qubits would grow to approach a steady state after a typical time depending on the parameters. An approximation to the steady state leaked population can be obtained from a simple three-level phenomenological model that we describe below.

The state in which a check qubit would be measured is modeled as being in one of three levels, the ground state (0), the excited state (1), or a leaked state (2, denoting any higher-energy quantum level). The probability distribution over the three possible states of each check qubit in cycle $i$ is defined by a column vector $\nu_i$, with the entries in the order of the states 0, 1 and 2.

Following one syndrome cycle, $\nu_i$ evolves to step $i+1$ according to a linear mapping, $ \nu_{i+1} = P\nu_i$,
and the transition matrix is,
\be  P= \left(\begin{array}{ccc}
 {1}/{2} + \delta &  (1-p_l)/2 + \delta & 0
 \\
 {1}/{2} - \delta &  (1-p_l)/2 - \delta & p_s
 \\
 0 & p_l & 1-p_s
   \end{array}\right),\ee
where we have defined for brevity $p_l\equiv p_{\rm leak},$ and $p_s\equiv p_{\rm seep}$. Here, $\delta$ is 0 without the flip protocol, which describes a scenario where check qubits have identical probabilities to be measured in 0 or 1 (due to the Pauli frame being random and ignoring biased readout errors), and qubits leak from (seep to) the level 1 with the leakage (seepage) probability. With $\delta \neq 0$, the model describes an increased qubit probability to be measured in the state 0, due to the flip protocol, with the exact magnitude of the phenomenological constant $\delta$ depending on other various error parameters of the model and decoding. 

$P$ is a right stochastic matrix, and we are interested in finding its steady state (the eigenvector corresponding to the eigenvalue 1). We consider the limit $p_l \ll p_s$, and in this limit we can write this eigenvector approximately as
\be v_{\infty} \approx \frac{1}{2}(1+2\delta, 1-2\delta, (1-2\delta)p_l/p_s).\ee

\begin{figure}
\centering
\includegraphics[width=0.48\textwidth]
{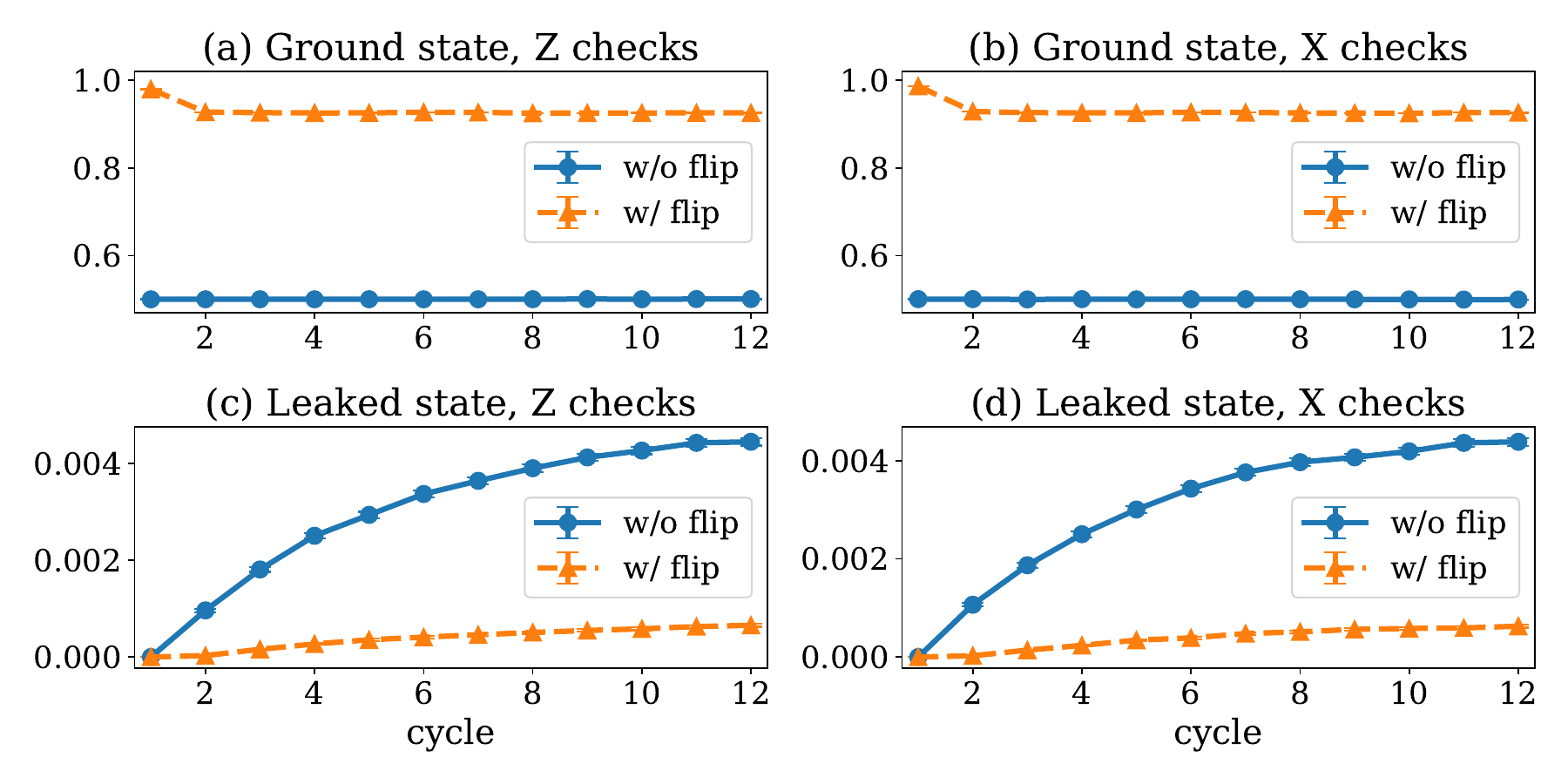}
\caption{The evolution of the mean number of leaked check qubits across the cycles of one syndrome measurement circuit, simulated using the state-dependent Pauli model as described above, with error rates as in \eqss{eq:fixed_rates2}{eq:medium_rates2} and $p_{\rm back}=0.025$. For the purpose of this figure, no qubits were initialized in a leaked state. (a), (b) The ground state population for Z and X checks. (c), (d) The leaked state population, for Z and X checks. Starting from an initial state with no leaked qubits, the number of leaked qubits grows and approaches a steady state. The mean population in the leaked state just before readout confirms the reduction of leakage by the flip protocol. 
\label{fig:evolution}}
\end{figure}

\begin{figure}
\centering
\includegraphics[width=0.47\textwidth]
{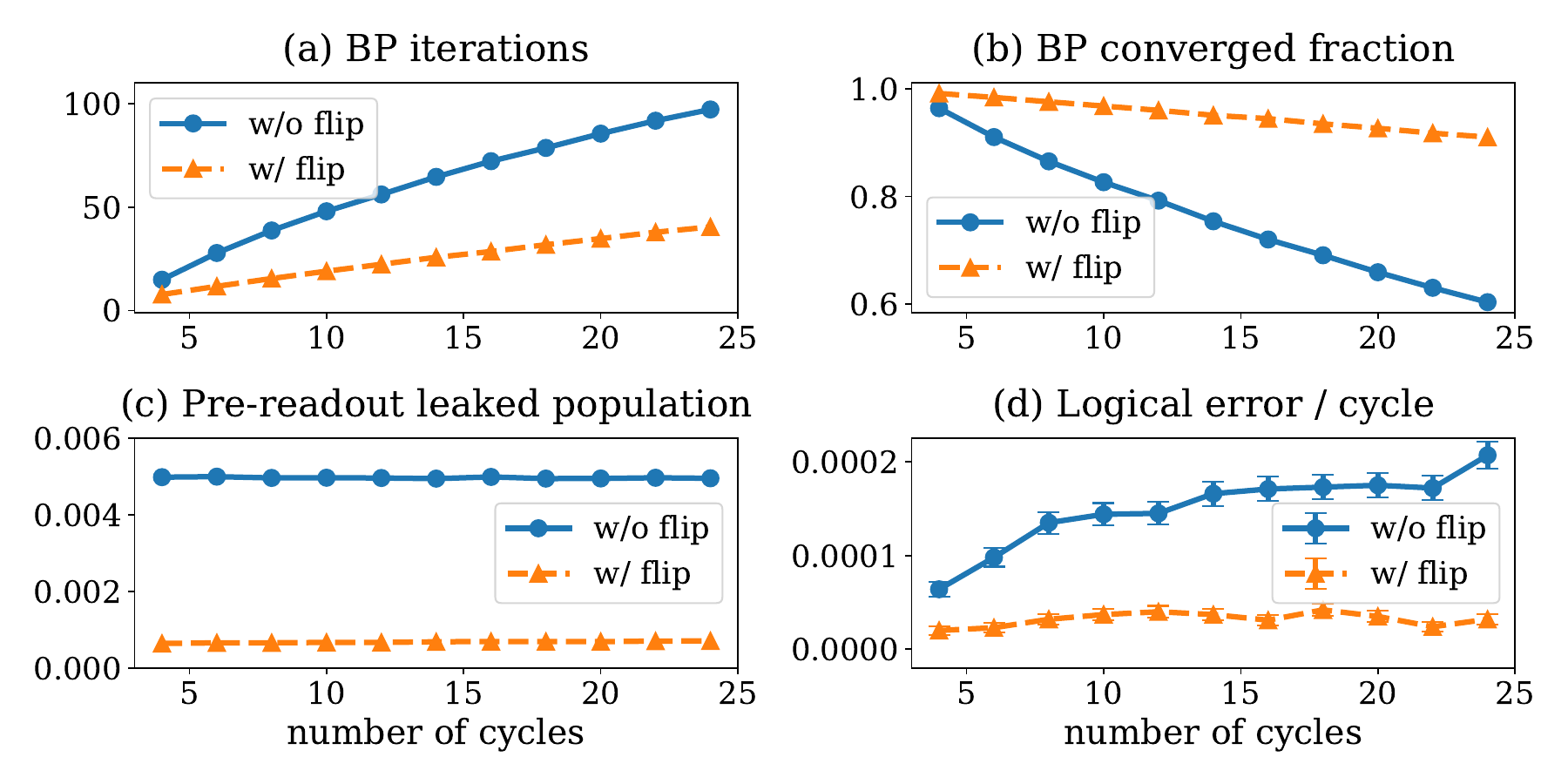}
\caption{The dependence of decoding characteristics on the number of cycles within the decoding window, for error parameters as in \fig{fig:evolution}. (a) The mean number of BP iterations per decoding window. (b) The fraction of BP algorithm runs that converged. (c) The mean population of leaked check qubits just prior to the syndrome cycles' readout step. (d) The logical error rate per cycle. We attribute the initial increase in the logical error to decoding faults that depend on the state and develop only after the first few cycles. 
\label{fig:cycles}}
\end{figure}

In \fig{fig:evolution} we show the evolution of the mean number of leaked check qubits across the cycles of one syndrome measurement circuit, simulated using the state-dependent Pauli model, when no check qubits are initially leaked. The error rates are set to
\be p_{\rm prep} = 0.001, \quad p_{\rm id,\,s} = 0.001, \quad p_{\rm cnot} = 0.001,\label{eq:fixed_rates2}\ee
the state-dependent readout parameters are,
\be p_{\rm read} = 0.02,\quad p_{\rm bias} = 0.005,\label{eq:readout_rates2}\ee
and we have also
\be p_{\rm id,\,l} = 0.005, \quad p_{\rm leak} = 0.002,\quad p_{\rm seep} = 0.2.\label{eq:medium_rates2}\ee
$\delta$ can be extracted numerically and is found to be roughly $\delta\approx 0.43$, giving a ground state population close to 0.93 and an almost one order of magnitude reduction in the leaked population of the checks. The rate of approach to the steady state is in this regime approximately $p_{\rm seep}$.
By initializing the state of the decoding simulation to a population of leaked qubits corresponding to the approximate steady state given above, we find that this is a very good self-consistent approximation, and the evolving state maintains a very similar population of the different levels (ground, excited and leaked). All simulations presented in the main text of this work are performed starting with a random qubit population with leakage.

We continue by looking at the dependence of the decoding quantities on the number of cycles within the decoding window. In order to properly account for readout errors in a code with distance $d$ it is a ``rule of thumb'' to decode about $d$ syndrome cycles together. However, this assumption deserves testing in a given error model and in particular with the state-dependent noise simulated as described above. In \fig{fig:cycles} we scan dependence of decoding characteristics for windows from 4 to 24 syndrome cycles. With the increase of the decoding window, the number of BP iterations increases and their converged fraction decreases, which could be expected due to the larger complexity of the decoding. However, this is accompanied with some increase that reaches an approximate plateau in the logical error per cycle, and we attribute this behavior to  decoding faults evolving from combinations of errors that depend on the state, and can develop only after the first few cycles. Since the decoding matrices grow linearly with the number of cycles and the actual time for running decoding simulations grows polynomially with the matrix sizes, running numerical simulations varying multiple parameters is computationally demanding. The question of optimizing the decoding is beyond the scope of the current work, and we take 12 cycles as an indicative good choice for all simulations presented here.

\bibliography{citations}

\end{document}